\documentclass[11pt]{article}
\usepackage[margin=1in]{geometry}
\usepackage{graphicx}
\usepackage{xspace}
\usepackage{amsmath}
\usepackage{multirow}

\def\Title#1{\begin{center} {\Large {\bf #1} } \end{center}}
\def\Author#1{\begin{center} {\normalsize {\sc #1} } \end{center}}
\def\Institution#1{\begin{center} {\normalsize {\it #1} } \end{center}}
\def\Abstract#1{\noindent {\normalsize {\bf Abstract:} {\normalfont #1}}}
\def\Conference{\vspace{4mm}\begin{raggedright} {\normalsize {\it Talk presented at the 2019 Meeting of the Division of Particles and Fields of the American Physical Society (DPF2019), July 29--August 2, 2019, Northeastern University, Boston, C1907293.} } \end{raggedright}\vspace{4mm}}

\newcommand{\fbinv}{\ensuremath{\text{fb}^{-1}}\xspace}
\newcommand{\pt}{\ensuremath{p_{\text{T}}}\xspace}
\newcommand{\ptgg}{\ensuremath{p_{\text{T}}(\gamma\gamma)}\xspace}
\newcommand{\ptg}{\ensuremath{p_{\text{T}}(\gamma)}\xspace}
\newcommand{\etag}{\ensuremath{\eta(\gamma)}\xspace}
\newcommand{\GeVc}{\ensuremath{\text{GeV}/\text{c}}\xspace}
\newcommand{\GG}{\ensuremath{\gamma\gamma}\xspace}
\newcommand{\mgg}{\ensuremath{m(\gamma\gamma)}\xspace}
\newcommand{\mH}{\ensuremath{m_{\text{H}}}\xspace}
\newcommand{\GeVcc}{\ensuremath{\text{GeV/c}^{2}}\xspace}
\newcommand{\MeVcc}{\ensuremath{\text{MeV/c}^{2}}\xspace}
\newcommand{\Hgg}{\ensuremath{\text{Higgs}\rightarrow\gamma\gamma}\xspace}
\newcommand{\Madgraph}{\ensuremath{\text{MadGraph}}\xspace}
\newcommand{\Pythia}{\ensuremath{\text{Pythia8}}\xspace}
\begin{document}

%
%

\Title{
Predictions for the Higgs boson mass measurement precision as a function of its transverse momentum up to 1 TeV for LHC and high luminosity LHC
}

\Author{Philip Baringer$^{(1)}$, Maxime~Gouzevitch$^{(2)}$, Anna Kropivnitskaya (speaker)$^{(1)}$}

\Institution{$^{(1)}$ University of Kansas, USA}
\Institution{$^{(2)}$ Universit\'{e} de Lyon, France}

\Abstract{
The question of naturalness of the Standard Model (SM) has been a hot topic since the discovery that the Higgs boson has a relatively light mass. It has been pointed out in the past that the mass of a scalar boson can be destabilized by loop corrections.
Many theories have been proposed beyond the SM to address this problem. It is possible that such mechanisms contribute to the running of the Higgs mass with the energy scale.
We present predictions for the precision of the Higgs mass measurement up to a Higgs boson transverse momentum of 1 TeV for the LHC in Run 3 with a luminosity of 300 \fbinv, and the high luminosity LHC with a luminosity of 3000 \fbinv. Predictions are generated with $\Madgraph$5, \Pythia and Delphes based on the CMS detector resolution.
}

\Conference

%
%

\section{Introduction}

In an effective-theory approach where momenta of virtual particles are cut off at the scale $\Lambda$, the quantum corrections to the physical Higgs mass grow proportionally with $\Lambda$~\cite{Pivovarov:2007dj,Giudice:2013nak}. In the one loop approximation for a fundamental Lagrangian the mass of the scalar particle can be written as:
\begin{equation}
\mH^2 = m_0^2+\Lambda^2\text{P}(\lambda_0,g)+\text{Q}^{2}\text{P}_1(\lambda_0,g)+\text{O}(\log{\text{Q}})\,,
\label{eq:mH}
\end{equation}
where $m_0$ is a parameter of the fundamental Lagrangian defined at the scale $\Lambda$, 
\mH is the ``measured" Higgs mass, P and $\text{P}_1$ are polynomials in the couplings, 
$\lambda_0$ is the Higgs self coupling, and $g$ is the dimensionless bare coupling of the model. 
There might in fact be a contribution from an intermediate energy scale $ \mH < \text{Q} << \Lambda$ that is usually neglected in the discussion. 
The size of this contribution may strongly depend on the exact beyond standard model (BSM) mechanism considered.

We can place the BSM models that address this question into two categories: those that consider 
$\mH \approx m_0$ and $\text{P}(\lambda_0,g) = 0$, like softly broken supersymmetry (SUSY)~\cite{Martin:1999hc}, and those that assume
$m_0^2 \approx -\Lambda^2\text{P}(\lambda_0,g) >> \mH$. The former class of model has been strongly ruled out by the 
latest LHC results~\cite{PhysRevD.98.030001}. 
In the latter category, a fine tuning of the 
parameters at high energy scale produces a low $\mH$ value at the electroweak symmetry breaking (EWSB) scale. In other words, following the 
definition of Ref. \cite{Williams:2015gxa}, in those theories we face the naturalness problem in the sense that there is a significant correlation between
the low EWSB scale and the high $\Lambda$ scale of the fundamental Lagrangian.
Unfortunately, direct searches at the LHC and in high precision lower energy data do not provide any clear hint regarding the nature of the underlying BSM theory.

In this work we propose an alternative approach to shedding light on the nature of the underlying BSM theory. The way that we usually look for BSM models at the LHC, we either  consider some direct evidence through a deviation in the production cross section of some phenomena, usually in high Q tails, or hunt for some rare or forbidden decays of known SM particles (H, Z, B-mesons, etc.).  In all cases the typical systematic and statistical uncertainties, amounting to several percent, are dominated by our estimates of the SM backgrounds and luminosity. Moreover, increased integrated luminosity usually does not bring relief since we face a flowering of the systematic uncertainties due to challenging effects. 

Naively, hadron colliders are considered as dirty machines that are not well equipped for precision measurements. Looking more deeply, however, we realize that there is an exception: the masses of the electroweak particles: W, Z, top, and Higgs. (For more detail, see the PDG review \cite{PhysRevD.98.030001}). The Z boson mass was measured with a 0.1 per mille precision by LEP collaborations and the LHC experiments are not competitive there. But the top and W masses were measured, respectively, with precisions of 0.3\% and 0.025\% at the LHC, which is very competitive with Tevatron collaborations, while the Higgs boson mass has been measured only at the LHC and with a 0.2\% precision \cite{Aad:2015zhl}. 

Why do we reach such a great precision on the measurement of the mass of an electroweak particle? The mass is extracted from the ``average value" of a peaking distribution or from a kinematic edge. The precision of the measurement evolves as $\sqrt{N_S}$, where $N_S$ is the number of the accumulated signal events and the only relevant systematic uncertainty is related to the energy calibration of the objects used in the analysis. The most precisely measured objects are photons $\gamma$ and leptons $l$ (electrons and muons). In this paper we consider the decay $H \rightarrow \gamma \gamma$, but one could also look at $H \rightarrow ZZ \rightarrow l^+l^-l^+l^-$. The contribution of a smooth background to this measurement is negligible provided that the signal significance exceeds 3 $\sigma$ and this is largely independent of the luminosity. 

Measuring $\mH$ with high precision as a function of transverse momentum (\pt) to constrain the dependence of $\mH$ on $Q$ in Equation~$\ref{eq:mH}$ is in fact an excellent goal for a high luminosity hadron collider such as the HL-LHC. It can provide insight with a sub-percent precision on the mechanism that generates low values of $\mH$ even for a value of $\Lambda$ that is well beyond the reach of the LHC.

\section{Analysis Strategy}

\subsection{Principle of the measurement}

We are looking for the production of a Higgs boson with  subsequent decay \Hgg. This is a well known golden channel, which was important to the discovery of the Higgs boson \cite{Chatrchyan:2012xdj} and for the simultaneous measurement of its mass. Despite the fact that the \Hgg branching fraction is very small (0.22\%), 
the fully reconstructed final state can be easily separated from the background by looking at the diphoton invariant mass (\mgg) distribution. The main idea behind the search is that the background produced by photon radiation from quarks, referred to as \GG QCD production, is smoothly evolving as function of \mgg, while the signal is a sharp peak with a resolution of 1-2 \GeVcc and centered around 125 \GeVcc. The electromagnetic calorimetry of the ATLAS and CMS experiments was optimized for the reconstruction of the photons in the energy range needed to find the Higgs boson in this channel.

In this analysis, signal extraction is performed by looking for a peak over the smooth background in the \mgg spectrum in bins of Higgs boson (or \GG) \pt. The Higgs \pt is used here as a proxy for the scale Q. The measured Higgs boson mass is taken to be the average value of the signal distribution obtained from a Gaussian fit.

\subsection{Simulation}

To determine the Higgs boson mass measurement precision as a function of \pt, we simulate gluon-gluon fusion production of \Hgg plus 0 and 1 jet from pp collisions at 13 TeV, generated in Higgs boson \pt bins. 
For the Higgs signal and \GG background simulations 
$\Madgraph\_$aMC@NLO~\cite{Alwall:2014hca} version 2.6.5 was used
with PDF NNPDF30\_nlo\_nf\_5\_pdfas (292200) and maxjetflavor = 5. 
Generated data were passed though \Pythia~\cite{Sjostrand:2014zea} fragmentation 
with the parameter QCUT - minimum kt jet measure between partons - set to 10.0 for the generator level and 15.0 for the fragmentation (\Pythia).
To simulate the detector response Delphes~\cite{deFavereau:2013fsa} was used with CMS resolution parameters for Run I.
The jet clustering procedure in Delphes was performed via the FastJet package~\cite{Cacciari:2011ma,hep-ph/0512210}.

To have a sufficient number of events at high \pt, it is necessary to generate the Higgs boson in several \pt bins.
The procedure is to generate a stable Higgs boson (not decaying)  in \Madgraph (pp$\rightarrow H$ and pp$\rightarrow H+$jet).
For example, to generate a Higgs in the \pt bin from 120 to 200 \GeVc 
the following parameters should be set in \Madgraph: 
\{25:120\} = pt\_min\_pdg and \{25:200\} = pt\_max\_pdg, where 25 is the Higgs particle identification number.
If the Higgs boson is decayed in \Madgraph, the Higgs \pt restriction isn't taken into account by \Madgraph.
That is why the decay \Hgg is done in \Pythia.
Due to parton showering effects, the Higgs \pt emerging from \Pythia could differ from the \pt in \Madgraph by several \GeVc, which is why the generated Higgs \pt bins were made wider than the bins used at the reconstructed level. 
In Table~\ref{tab:MadgraphHiggsBins}, the generated and reconstructed Higgs and \GG \pt bins are presented. The two last reconstructed bins are combined for the analysis, due to the small reconstructed Higgs rate for the HL-LHC.

The dominant background is QCD prompt diphoton production. Additional backgrounds, amounting to roughly 20\% of the total, come from so-called \textit{fake photons}, \textit{ie.}, jets that have been misidentified as photons. These fakes mainly arise from the decay of a leading $\pi^0$ in a jet, a quite rare situation that may occur in the tail of fragmentation functions. One of the main tasks of the CMS and ATLAS photon identification is to separate the prompt $\gamma$ from the in-jet $\pi^0\rightarrow \GG$ decays using shower shapes and isolation variables. This level of detail is hard to simulate and it is not well emulated by Delphes. It happens that the \mgg distribution for $\gamma$-jet and jet-jet contributions is similar in shape to the \GG contribution, so we vary the normalization of latter to account for the former \cite{Sirunyan:2018ouh}.

\begin{table}[t]
\begin{center}
\begin{tabular}{l|cccccccc} \hline 
Level & \multicolumn{8}{c}{Binning in \pt [\GeVc]}
\\ \hline\hline
\Madgraph \\ Generated Higgs &   
0-inf & 40-250 & 110-350 & 150-450 & 200-550 & 300-650 & 380-900 & 580-inf 
\\ \hline
\Madgraph \\ Generated \GG &   
0-inf & 70-250 & 140-350 & 180-450 & 250-550 & 310-650 & 380-900 & 580-inf 
\\ \hline
Reconstructed &
0-120 & 120-200 & 200-270 & 270-350 &  350-450 & 450-550 & 550-750 & 750-inf
\\ \hline
Reconstructed \\ used in analysis&
0-120 & 120-200 & 200-270 & 270-350 &  350-450 & 450-550 &\multicolumn{2}{c}{550-inf}
\\ \hline
\end{tabular}
\caption{Generated Higgs and \GG \pt bins are presented in the first and second row, respectively.
Reconstructed-level \pt bins are presented in the third row.
The two last reconstructed bins are combined for the analysis, due to the small reconstructed Higgs rate for the HL-LHC, as presented in the last row.}
\label{tab:MadgraphHiggsBins}
\end{center}
\end{table}

\subsection{Selection}
\label{sec:Selection}

The invariant mass of the \GG system, \mgg, is reconstructed to find the Higgs boson signal. 
The following event selection criteria are applied, which follows the CMS selections for the 2016 data~\cite{Sirunyan:2018ouh}:
\begin{itemize}
\item Barrel photon (B): $|\etag| < 1.44$;
\item Endcap photon (E): $1.57 < |\etag| < 2.5$;
\item Select \GG pairs that are either barrel-barrel (BB) or barrel-endcap (BE);
\item Leading photon      $\ptg > \mgg/3$;
\item Subleading photon $\ptg > \mgg/4$;
\end{itemize}
In addition we request a leading jet with $\pt > 30$ GeV. This condition was added to emulate the gluon-gluon fusion selection with a hard recoiling jet. (In future studies this requirement might be removed since it impacts significantly the sensitivity to \mH at low \pt.)

\begin{figure}[htb]
\centering
\includegraphics[height=8.05cm]{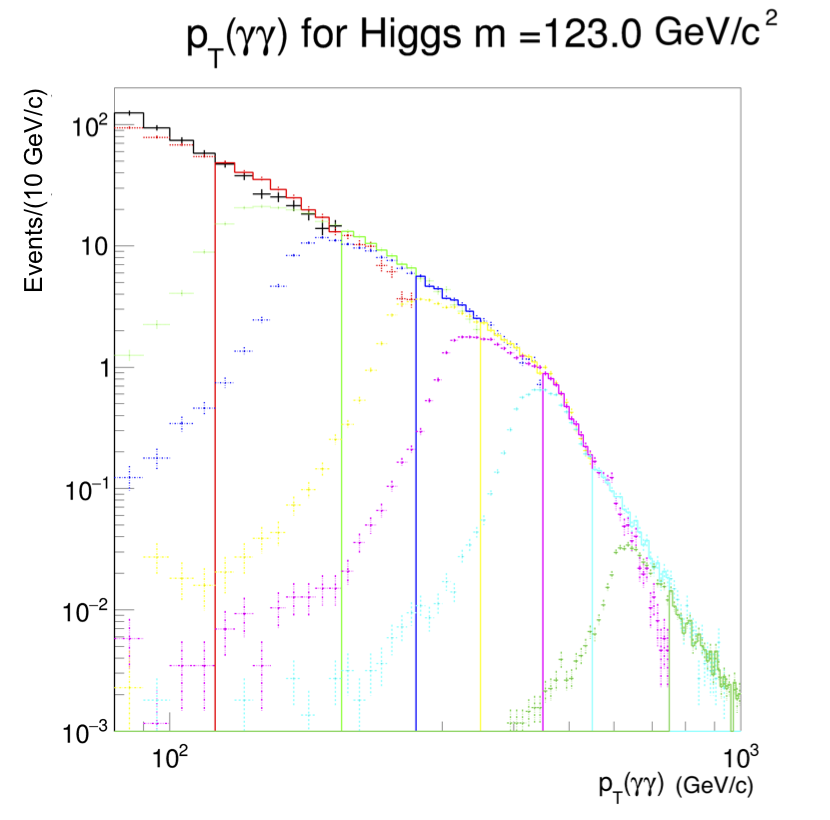}
\includegraphics[height=8.05cm]{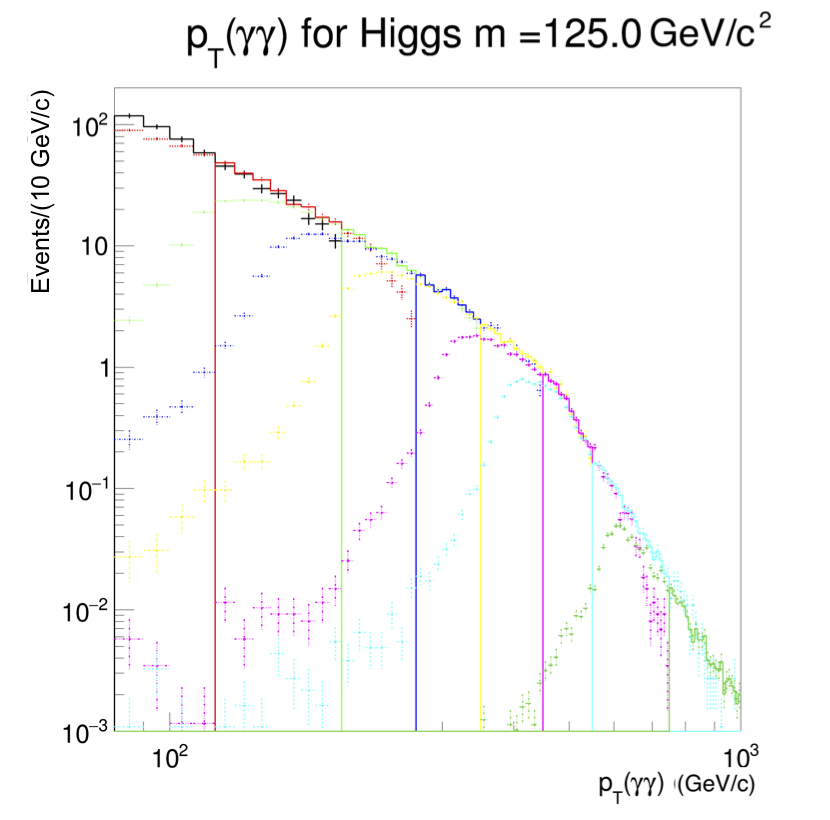}\\
\includegraphics[height=8.05cm]{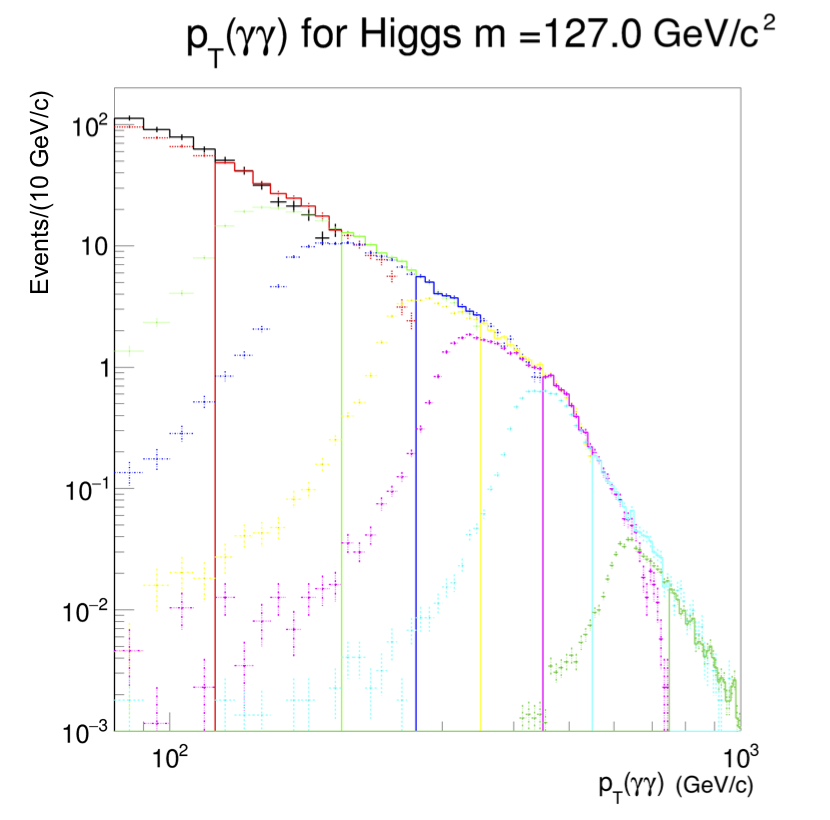}
\includegraphics[height=8.05cm]{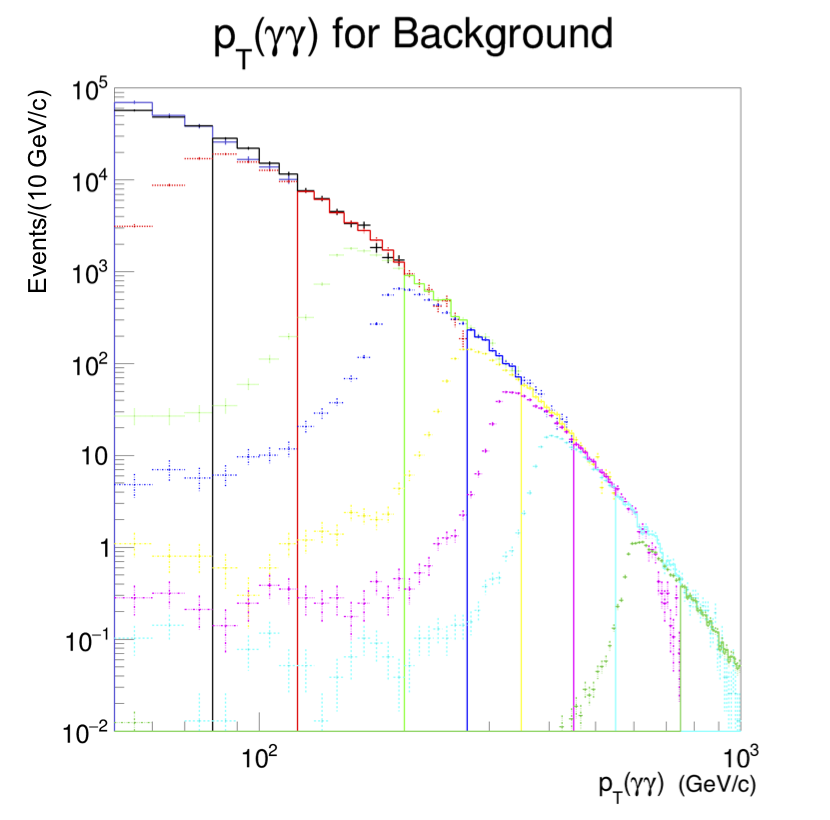}
\caption{Results for the \ptgg distribution after the reweighting process
for Higgs mass 123 (top-left), 125 (top-right) and 127 (bottom-left) \GeVcc
and \GG background (bottom-right).
Solid lines correspond to reconstructed bins, 
while dashed lines correspond to the reconstructed spectrum after generation process.}
\label{fig:PtMerge}
\end{figure}

Signals with Higgs masses of 123, 125 and 127 \GeVcc, and the \GG background are generated 
in eight bins of Higgs \pt   and \ptgg, with 50000 events in each sample as shown in Table~\ref{tab:MadgraphHiggsBins}. 
Bins were reweighted to the generated cross section. 
Results for the \ptgg distribution after the reweighting process are presented at Fig.~\ref{fig:PtMerge}.

\begin{figure}[htb]
\centering
\includegraphics[height=12cm]{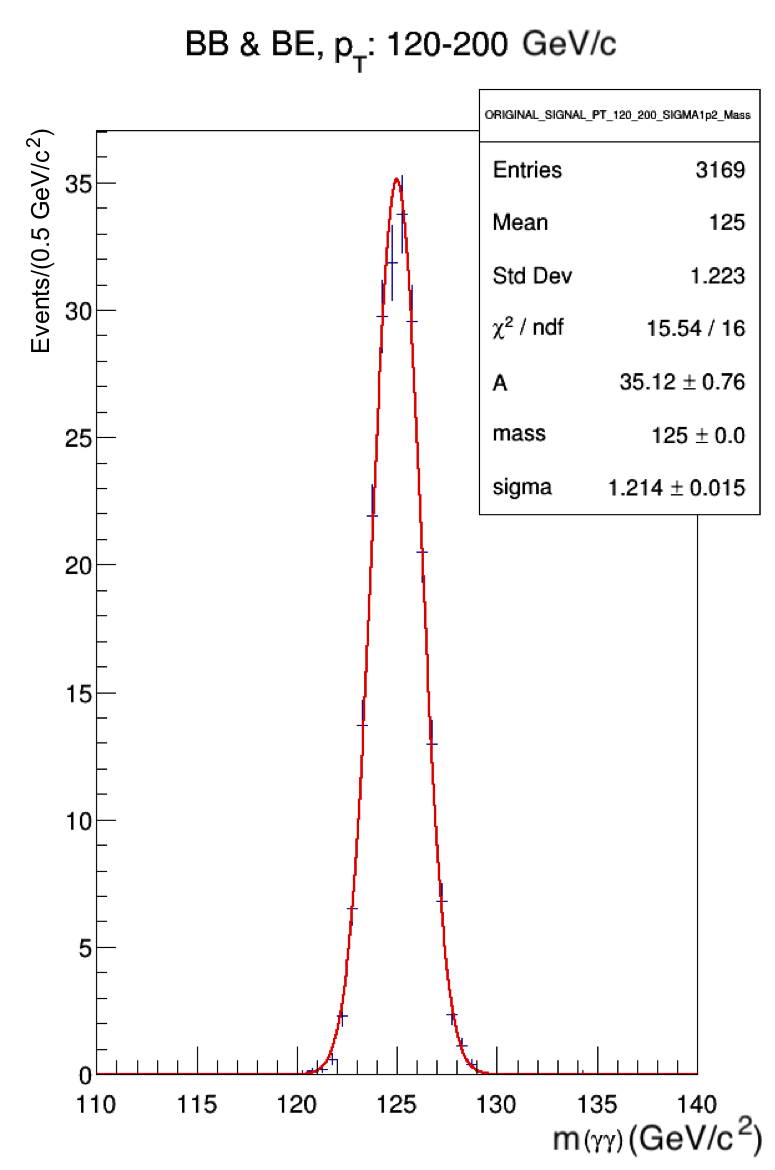}
\includegraphics[height=12cm]{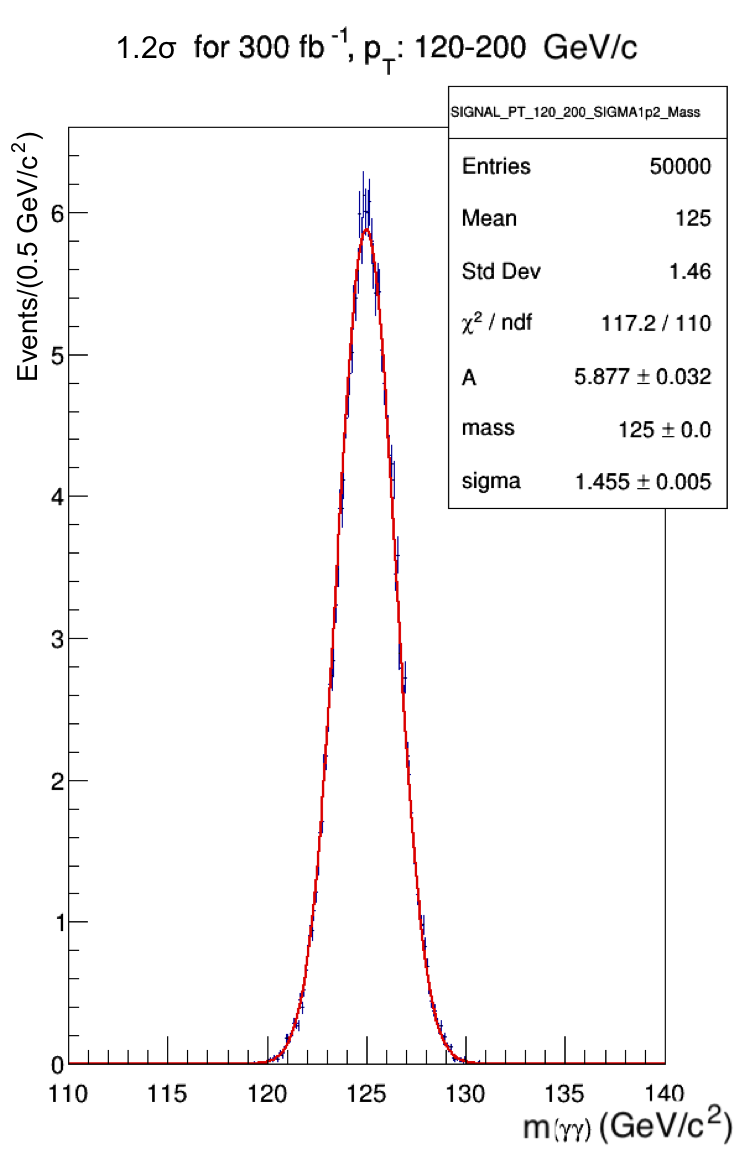}
\caption{The simulated Higgs signal after Delphes reconstruction in the BB \& BE regions is presented for the Higgs \pt bin from 120 to 200 \GeVc (left). 
Scaled Higgs signal by regenerating with wider Gaussian shape (1.2$\sigma$), smaller binning 
(from 500 \MeVcc to 100 \MeVcc) and higher statistics,  normalized to 300 \fbinv (right).
The red line in each plot corresponds to the Gaussian fit.}
\label{fig:signal}
\end{figure}

In real CMS data the barrel photons are much better reconstructed than the endcap ones because of the lower amount of material in front of the electromagnetic calorimeter in the barrel. Therefore the barrel-barrel (BB) diphoton mass resolution is usually better than the barrel-endcap (BE) resolution. But the Delphes CMS Run I data cards used here do not emulate those details, therefore we can combine the BB and BE
regions into one sample ``BB \& BE". 
The resulting \mgg distribution for the Higgs signal could then be fit using a simple Gaussian distribution. 
In Fig.~\ref{fig:signal} (left), the simulated Higgs signal is presented after Delphes reconstruction in the BB \& BE regions with a Gaussian fit (red line) for the Higgs \pt bin from 120 to 200 \GeVc. 
The Higgs mass resolution is around 20\% better than in the CMS paper, so the signal was regenerated using the Gaussian shape from the fit, but with the width increased by 20\% (1.2$\sigma$). The new shape was generated with smaller binning (going from a 500 \MeVcc bin width to 100 \MeVcc) and higher statistics (50000 entries per Higgs \pt bin), but keeping the normalization appropriate to a data sample of 300 \fbinv.
The rescaled distribution is presented in Fig.~\ref{fig:signal} (right).

\begin{figure}[htb]
\centering
\includegraphics[height=8.2cm]{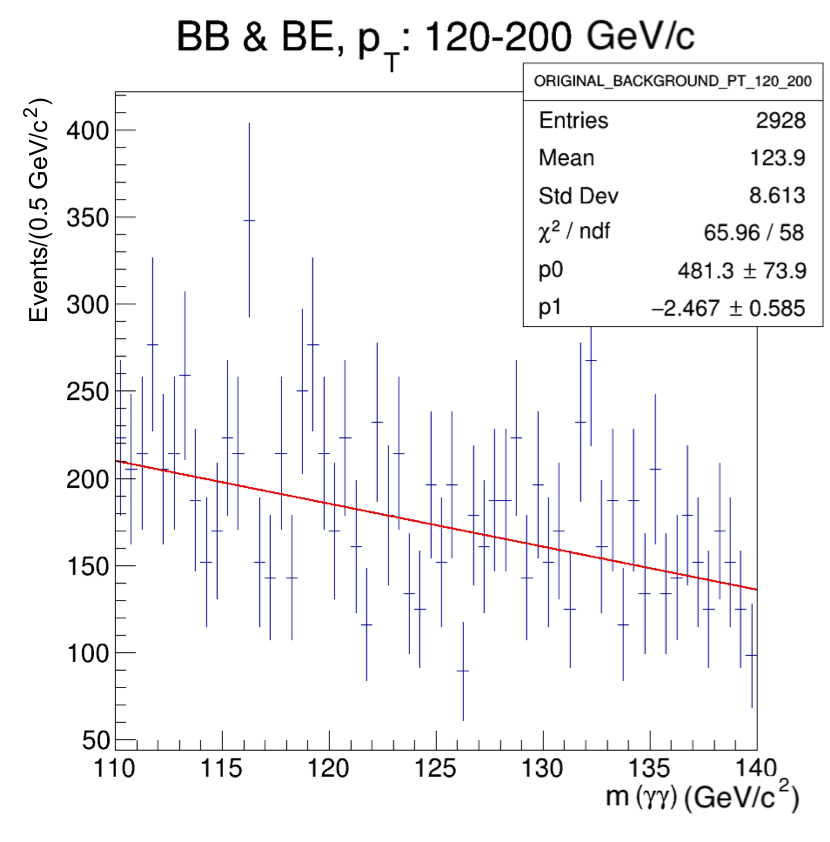}
\includegraphics[height=8.2cm]{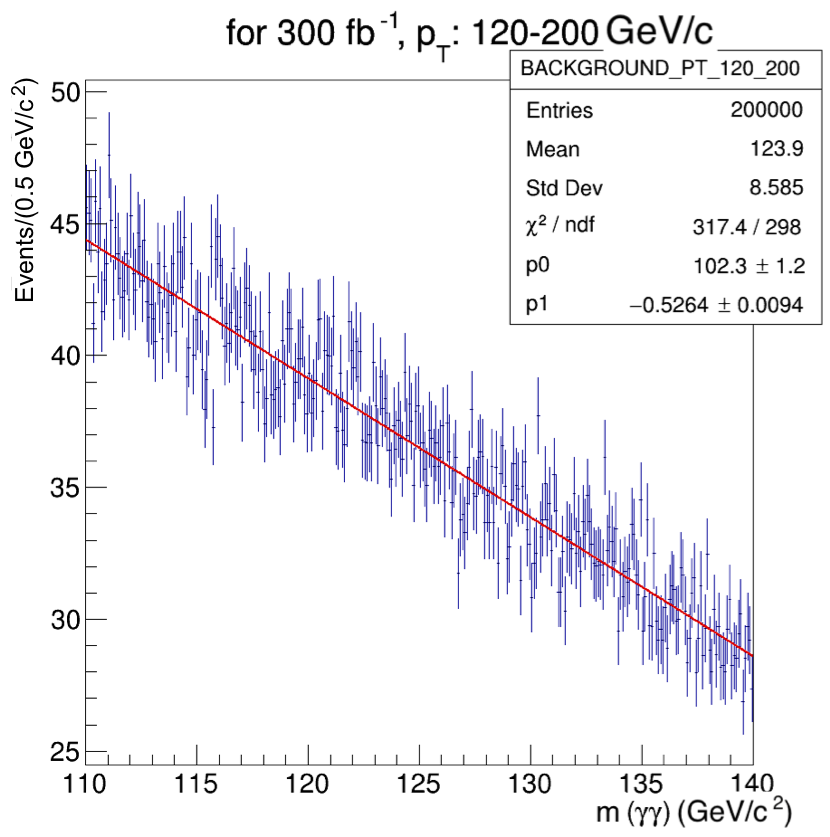}
\caption{The reconstructed \GG background after Delphes reconstruction in the BB \& BE regions is presented for the \ptgg bin from 120 to 200 \GeVc (left).
Scaled background 
after regenerating with smaller binning (from 500 \MeVcc to 100 \MeVcc) and higher statistics, normalized to 300 \fbinv (right).
The red line in each plot corresponds to the linear fit.}
\label{fig:background}
\end{figure}

In Fig.~\ref{fig:background} (left), the simulated \GG background is presented
after Delphes reconstruction in the BB \& BE regions with a linear fit (red line) 
for the \ptgg bin from 120 to 200 \GeVc.
As was done with the signal, the background was regenerated with the (linear) shape taken from the fit and with smaller binning (going from a 500 \MeVcc bin width to 100 \MeVcc) and higher statistics (50000 entries per \ptgg bin), but no resolution correction is applied in this case. The distribution, normalized to 300 \fbinv, is presented in Fig.~\ref{fig:background} (right).

\begin{figure}[htb]
\centering
\includegraphics[height=5cm]{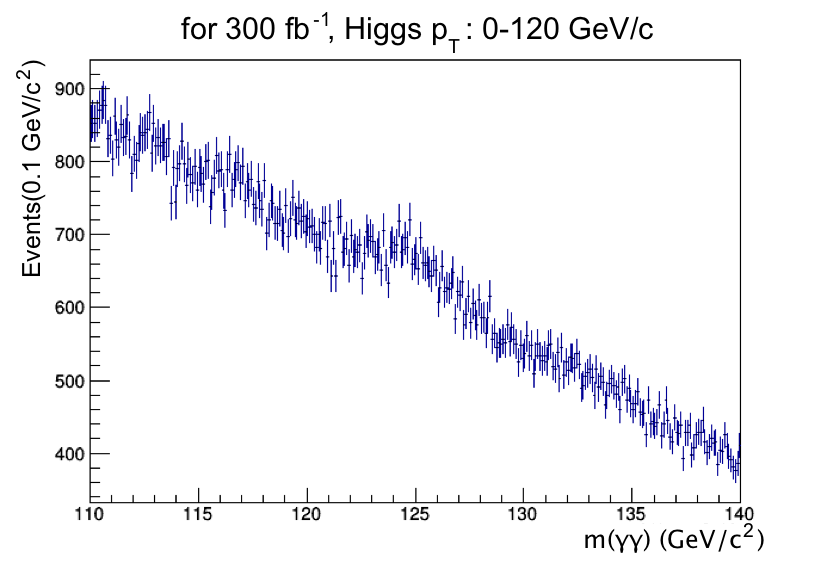}
\includegraphics[height=5cm]{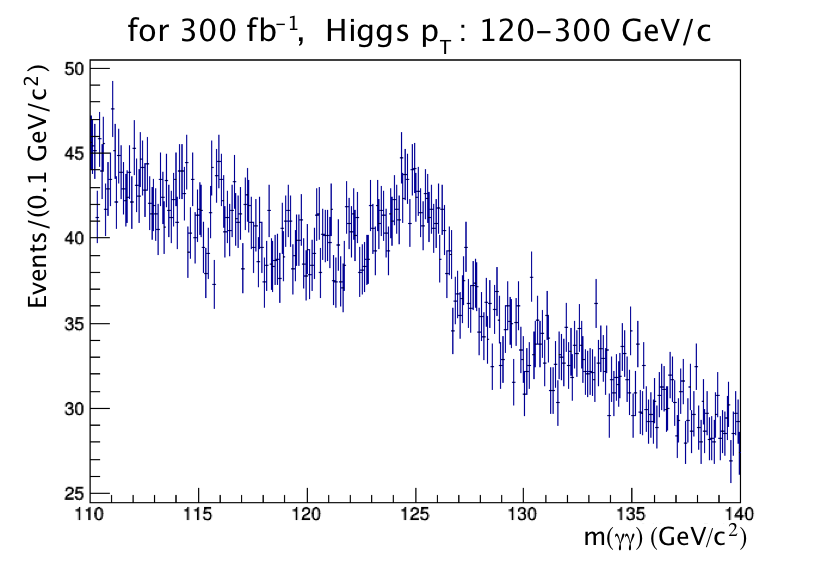}\\
\includegraphics[height=5cm]{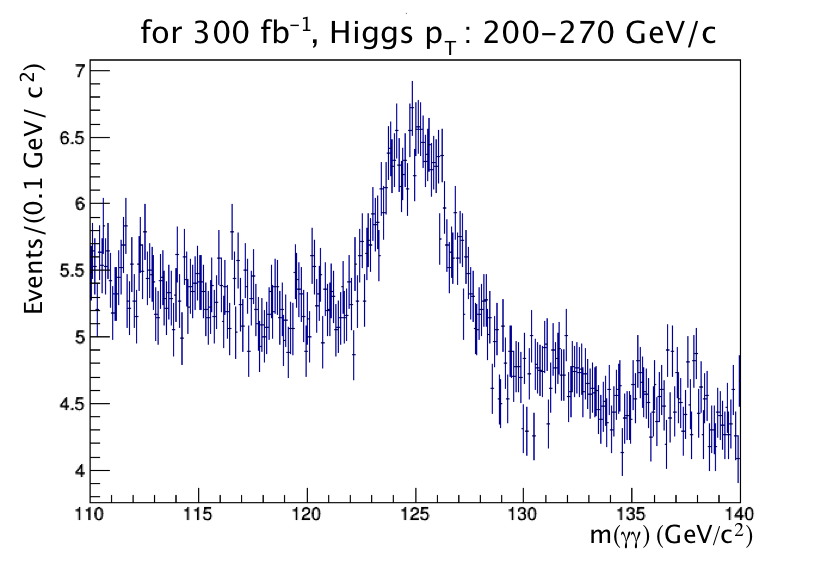}
\includegraphics[height=5cm]{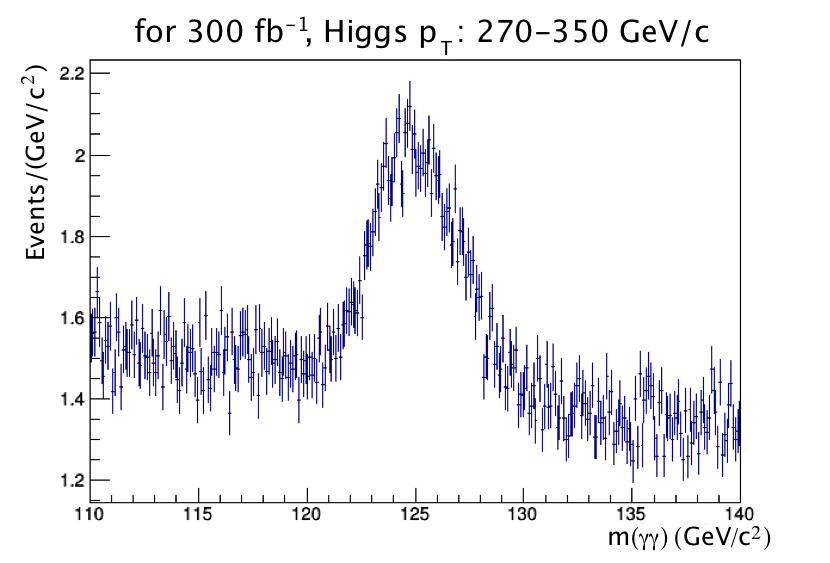}\\
\includegraphics[height=5cm]{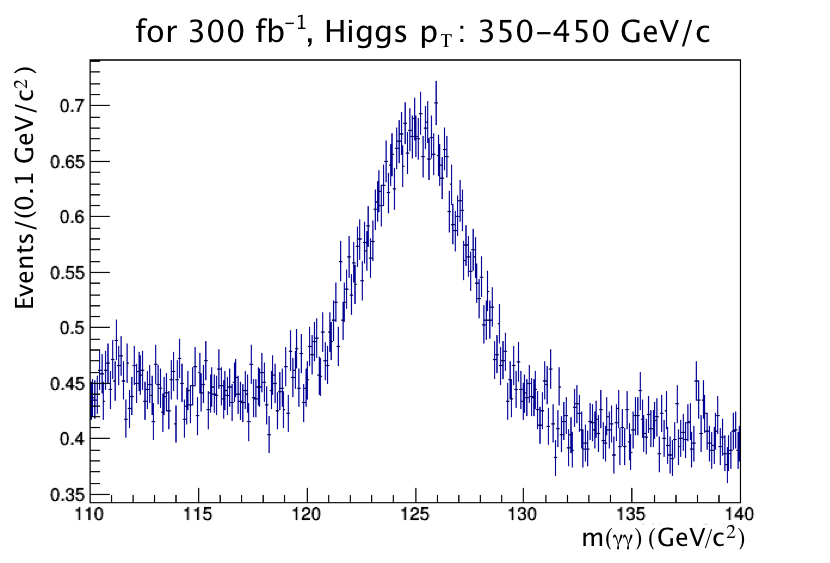}
\includegraphics[height=5cm]{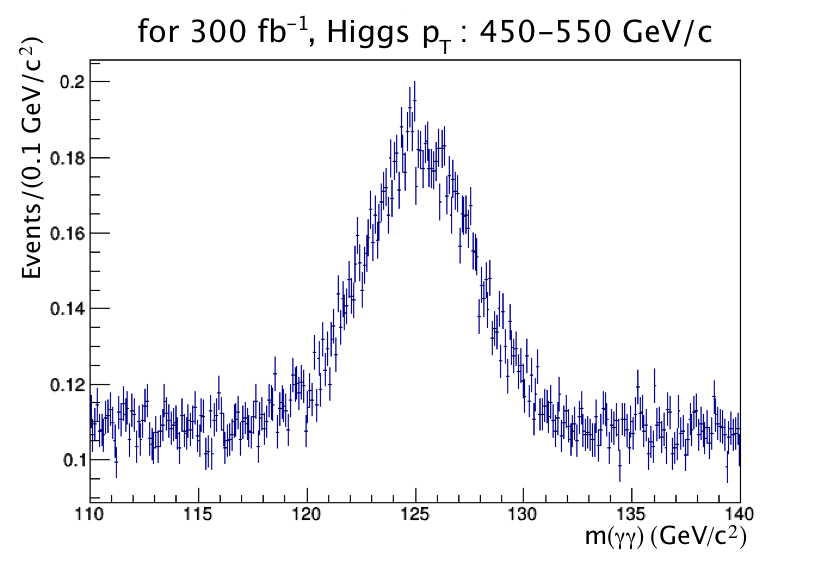}\\
\includegraphics[height=5cm]{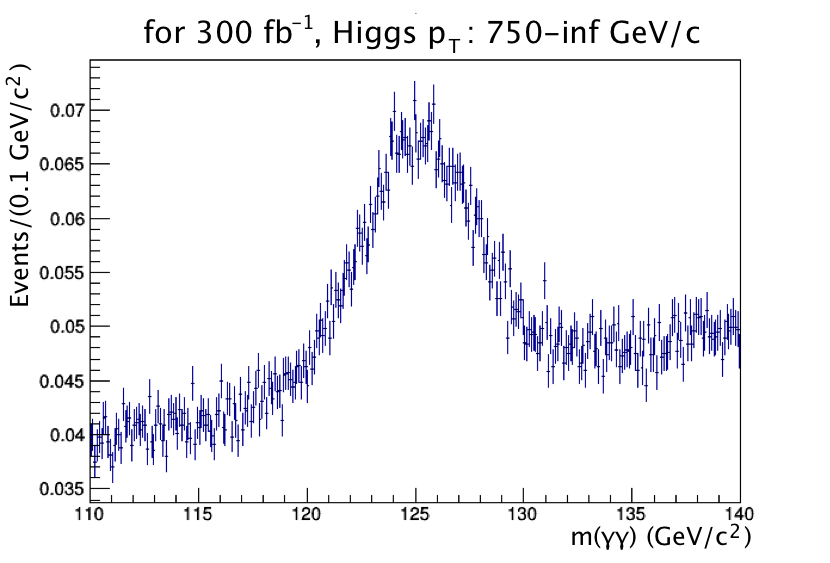}
\caption{The rescaled signal over the background for different Higgs \pt bins.}
\label{fig:SandB}
\end{figure}

The rescaled signals and backgrounds are used in the determinations of the Higgs mass measurement precisions presented in Section~\ref{sec:SignalExtraction}. Their sum for the luminosity of 300 \fbinv is shown in Fig.~\ref{fig:SandB}.

\subsection{Signal extraction procedure}
\label{sec:SignalExtraction}

\begin{figure}[htb]
\centering
\includegraphics[height=8.1cm]{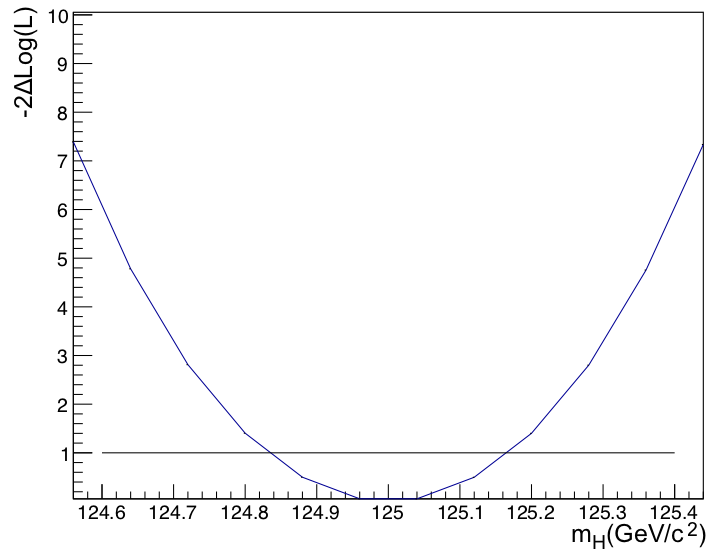}
\caption{The normalized likelihood scan of the \mH for Higgs \pt bin from 120 to 200 \GeVc.
}
\label{fig:Likelihood}
\end{figure}

For the mass estimation, the CMS Higgs Combine tool is used~\cite{ATLAS:2011tau,Conway:2011in}
and the following strategy is applied:
\begin{itemize}
\item Build statistical binned likelihood.
\item Inject 125 \GeVcc standard model Higgs signal.
\item Scan likelihood as a function of the Higgs mass hypothesis.
\item Use templates interpolated from the 123, 125, and 127 \GeVcc simulations.
\end{itemize}
The Higgs mass precision was estimated in the seven Higgs \pt bins listed in Table~\ref{tab:MadgraphHiggsBins} (last row).
In Figure~\ref{fig:Likelihood}, the normalized likelihood scan of \mH for the Higgs \pt bin from 120 to 200 \GeVc is presented.

\section{Results}

In Table~\ref{tab:HiggsRate}, the expected number of Higgs signal and background events
for an HL-LHC data sample corresponding to 3000 \fbinv at a center-of-mass energy of 14 TeV is presented.
To make predictions for the HL-LHC, the 13 TeV Higgs simulation was scaled to the 14 TeV Madgraph prediction using the theoretical cross sections: the scale factor varies from 1.14 to 1.29 from the lowest to the highest Higgs \pt bin.

\begin{table}[t]
\begin{center}
\begin{tabular}{l|ccccccc} \hline 
Higgs \pt [\GeVc]&  0-120 &  120-200 &  200-270 & 270-350 & 350-450 & 450-550 & 550-INF
\\ \hline \hline
Higgs  &   
16 100 & 2 500 & 1 200 & 350 & 180 & 61 &20
\\ \hline
Background\\ 
$120<\mgg<130$ 
\\ \GeVcc &
469 000 & 31 000 & 4 600 & 1 360 & 410 & 110 & 48
\\ \hline
\end{tabular}
\caption{Expected number of Higgs signal and background events for an HL-LHC data sample of 3000 \fbinv (14 TeV).}
\label{tab:HiggsRate}
\end{center}
\end{table}

\begin{figure}[htb]
\centering
\includegraphics[height=8.cm]{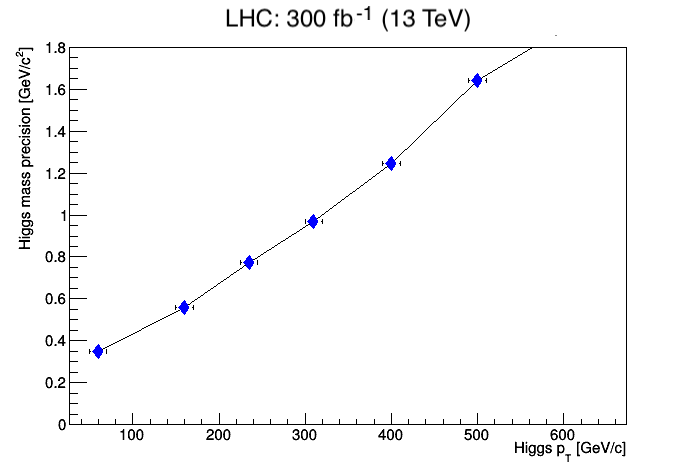}\\
\includegraphics[height=8.cm]{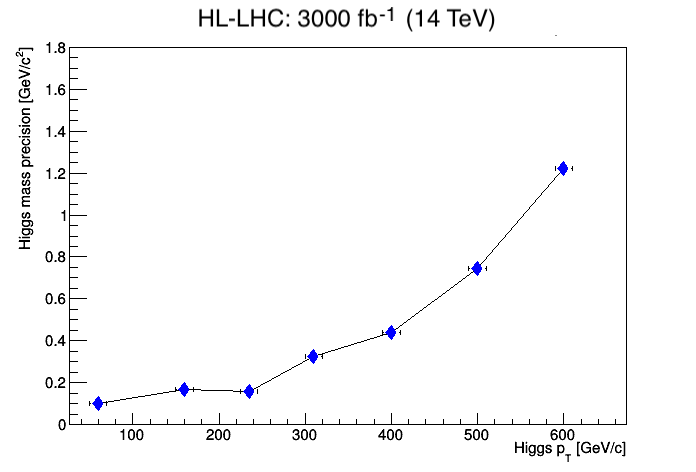}
\caption{First results for expected Higgs mass precision, without systematic uncertainties, for LHC (300 \fbinv) (top) and HL-LHC (3000 \fbinv) (bottom).}
\label{fig:MassResolution}
\end{figure}

In Figure~\ref{fig:MassResolution}, the first results for the expected 
Higgs mass precision, without an estimate of systematic uncertainties, for the LHC with a luminosity of 
300 \fbinv (top) 
and the HL-LHC with 3000 \fbinv (bottom) are presented.

\begin{figure}[htb]
\centering
\includegraphics[height=8.cm]{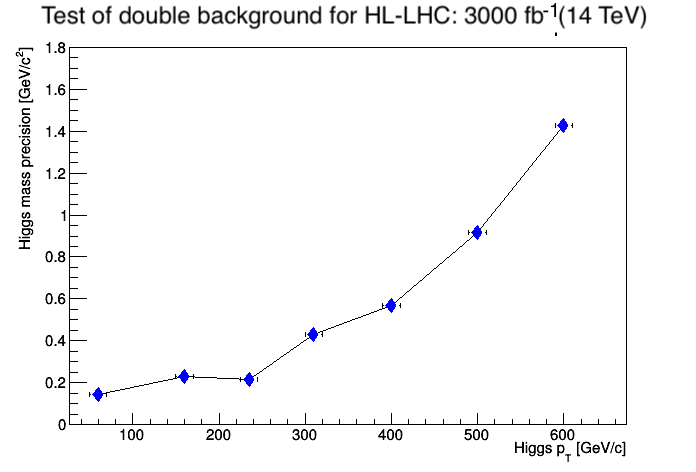}
\caption{Test for expected Higgs mass precision for HL-LHC (3000 \fbinv) by increasing \GG background by a factor of two.}
\label{fig:MassResolution_2bg}
\end{figure}

Currently, no energy scaling is applied to the background. 
In Figure~\ref{fig:MassResolution_2bg}, the effect of a larger background due to fake photons and other factors was explored by increasing the \GG background by a factor of two:
the uncertainty on the mass measurement increases by less than 30\% for Higgs $\pt>200\GeVc$.
This variation covers potential background contributions from $\gamma$-jet and jet-jet processes.  

A significant drop in the \Hgg reconstruction efficiency is observed in the current generated sample. 
For example, for the last Higgs \pt bin used in the analysis (Higgs $\pt > 550$ \GeVc),
the efficiency drops to 20\%. This effect comes from the photon isolation criteria used in the Delphes detector simulation, 
which is around 0.5 for $\Delta R =\sqrt{\Delta\eta^2+\Delta\phi^2}$. 
If $\Delta R$(\GG) is below 0.5, then the Higgs is reconstructed 
as a fat jet in Delphes. This fat jet contribution becomes significant starting from a Higgs $\pt$ of about 270 \GeVc. 
The contribution from simple jets starts from a Higgs \pt of around 450 \GeVc.
Thus, the results in this paper are conservative estimates due to our low efficiency in the high \pt bins. 
Improvement is expected in future studies by correcting the isolation criteria in Delphes.


\section{Conclusions}

The first preliminary results are presented for high-\pt Higgs mass measurement resolutions for 300 \fbinv at the LHC and 3000 \fbinv without systematic uncertainty estimation at the HL-LHC 
for a CMS-like detector. 
In run 3 at the LHC, we could be sensitive to mass differences of 1.2 \GeVcc 
up to a Higgs \pt of 350-450 \GeVc. At the HL-LHC, we could reach the same mass precision for a Higgs \pt above 550 GeV.

We plan to repeat this analysis for the HL-LHC Delphes configuration built for the CERN Yellow Report for European Strategy I 2019, improve the photon isolation criteria in Delphes,
estimate the contribution to the background from $\gamma$-jet and jet-jet diphoton fakes, 
and estimate systematic uncertainties. 
We are expecting some improvement in sensitivity over the results reported here. 

\section*{Acknowledgements}
We are grateful to support from the National Science Foundation.
Motivation for this measurement was built in collaboration with Grigorii Pivovarov and Victor Kim. The Masters student Manon Bourgade, and undergraduate students Nicole Johnson and Eilish Gibson also contributed to the early stages of this work.

\end{document}